\documentstyle[11pt,aaspp4]{article}

\lefthead{Madore \& Freedman}
\righthead{Cepheid Distance Scale}

\begin{document}

\title{HIPPARCOS PARALLAXES\\AND\\
    THE CEPHEID DISTANCE SCALE}

\author{Barry F. Madore}
\affil{NASA/IPAC Extragalactic Database, California Institute of Technology,
    Pasadena, CA~~91125}

\and

\author{Wendy L. Freedman}
\affil{Observatories, Carnegie Institution of Washington,
    813 Santa Barbara Street, Pasadena, CA~~~91101}

\begin{abstract}
Hipparcos  parallaxes have recently become  available  for a sample of
Galactic Cepheids,  and we have used  these new distances to calibrate
the  Cepheid period-luminosity   (PL)   relation at six    wavelengths
($BVIJHK$).  Comparing these   calibrations with previously  published
multiwavelength  PL      relations  we   find   agreement  to   within
$0.07~\pm0.14$~mag,   or 4~$\pm7$\% in   distance.  Unfortunately, the
current parallax  errors  for the  fundamental pulsators  (ranging  in
signal-to-noise =  $\pi   /\sigma_{\pi}$ from 0.3   to 5.3,  at  best)
preclude  an  unambiguous interpretation of the  observed differences,
which may arise from a combination of true distance modulus, reddening
and/or  metallicity effects.  We  explore   these effects and  discuss
their implications for the distance to the  Large Magellanic Cloud (LMC) 
and the Cepheid-based extragalactic distance  scale. These results suggest
a range of LMC moduli between 18.44 $\pm$0.35 and 18.57 $\pm$0.11~mag;  
however,  other effects on the Cepheid PL relation ({\it  e.g.,} extinction, 
metallicity, statistical errors) are still as significant as any such 
reassessment of its zero point.
\end{abstract}


\keywords{Cepheids -- Magellanic Clouds -- infrared: stars}

\vfill\eject
\section{Introduction}

Feast  and Catchpole (1997 =  FC97  hereafter) have recently published
the  first  results  on  parallaxes   to Galactic   Cepheids  based on
measurements from the Hipparcos satellite.  They list  data for the 26
highest  signal-to-noise Cepheid parallaxes;  and  after an  extensive
series of reductions (see  their Table 2)  they conclude that the best
fit PL relation for the  visual bandpass is $M_V~=~-2.81~log(P)~-~1.43
$, with a standard error on the Hipparcos zero point of $\pm0.10$~mag,
adopting the slope from prior work on LMC Cepheids.  The authors go on
to  apply this V-band solution to  determining the distance modulus of
the  LMC corrected   for $E(B-V)~=~0.074$~mag.   Adding  a metallicity
correction of +0.042~mag and adopting $<V>_o-~log(P)$ from Caldwell \&
Laney (1991) gives  $(m-M_V)_o^{LMC}~=~18.70\pm0.10$~mag. In this {\it
Letter}  we  go  beyond    the  V-band PL  relation  and   explore the
implications of  the   Hipparcos    data   for   the   multiwavelength
calibrations  of the Cepheid  PL relation  from  the blue (B-band) out
into the near infrared (2.2$\mu$m K-band).

\section{Comparison with V-Band Period-Luminosity Relations}

In Figure~1 we compare  differentially four calibrations (heavy dotted
lines) of the V-band Cepheid PL relation with the FC97 Hipparcos-based
relation (solid horizontal lines).   The first two comparisons (in the
upper two panels) are  with the relations  given by Madore \& Freedman
(1991;    hereafter MF91), derived  from self-consistent   sets of LMC
Cepheid data   whose  stars either had complete    $BVRI$ observations
(MF91.1  containing 32  Cepheids)  or  complete $BVRIJHK$ observations
(MF91.2 containing 25 stars)\footnote{Tanvir (1997) has pointed out that 
there may be a small correction  amounting to 0.04~mag to  the published 
I-band magnitudes of these LMC Cepheids due to the originally sparse 
sampling and consequent averaging of their light and color curves. 
For the past five years we have been 
obtaining new VI CCD observations of the LMC calibrators at Las Campanas 
and now also at Siding Springs Observatories. These new data are designed 
to address those concerns.}   These first two solutions indicate the
sensitivity of slopes  and zero points  to sample selection, which are
considerable,  but  within   the  quoted  statistical   uncertainties:
$\pm$0.11 and  $\pm$0.20, respectively for  the  slopes, and $\pm$0.05
and $\pm$0.09~mag, for the zero points.  So as  to make the subsequent
comparisons  consistent,   the  original  Sandage   \&  Tammann (1968)
calibration (ST68.1 in  the lower left  panel) has been  placed on the
modern  Hyades/Pleiades Galactic cluster  distance scale by applying a
single offset of +0.13~mag derived from the average difference between
the absolute magnitudes  of the Cepheids used  in the 1968 calibration
updated  to Feast  and Walker (1987),  their Table  2.   This distance
scale corresponds to a Hyades modulus of  3.27 (see Pel 1985) and uses
the Pleiades main sequence, at  a modulus of  5.57 (van Leeuwen, 1983)
to effectively   correct for the over-metallicity   of the Hyades with
respect   to the older    Galactic   clusters  in which   the  Cepheid
calibrators  are found.\footnote{At the  February 14,  1997 meeting of
the Royal Astronomical Society in London on  February 14, 1997 F.  van
Leeuwen and C.S.  Hansen Ruiz reported a true distance modulus of 5.29
$\pm$     0.06~mag for the    Pleiades   cluster,  based on  Hipparcos
trigonometric parallaxes. Following the Venice Meeting in June 1997 the 
value had changed only slightly to 5.33 $\pm$0.06~mag (C. Turon, private 
communication)  If adopted, this  Pleiades modulus would make  the
Galactic-cluster-based calibrations approximately 0.3~mag fainter than
the FC97 solution plotted in Figure~1. At this point in time the 
Galactic cluster zero point appears to be in a state of flux, and we will 
not comment on it further, except to note that the Hipparcos calibration 
will undoubtedly converge on a more accurate zero point than we have 
access to at this precise moment.}  Finally, the FW87 calibration
itself is plotted in the  lower left panel.   In all panels the dashed
horizontal lines represent  the fiducial Hipparcos calibration flanked
by thin parallel lines at $\pm0.10$~mag.


\placefigure{fig1}

The error  bars of all of  the plotted previously  published relations
overlap  with  errors  quoted for  the  Hipparcos solution  (a  formal
uncertainty was not  given by ST68,  so  we have  arbitrarily assigned
them  an  error  of  $\pm0.05$~mag).   However,  the  offsets  are not
randomly distributed, with   each of  the  solutions appearing  to  be
systematically fainter in      $V$ with  respect  to   the   Hipparcos
calibration by   about  0.1~mag.   We  discuss the   significance  and
possible implications of this difference in the following sections.

\section{Multiwavelength Period-Luminosity Relations}

In  Madore \& Freedman  (1991) we  published fiducial  PL relations in
seven  bandpasses: $BVRIJHK$.    These  were all  based on   selecting
self-consistent sets of  previously published LMC Cepheid data, scaled
to  an LMC true distance modulus   of 18.50~mag and  applying a single
line-of-sight  reddening   correction   using   $E(B-V) =   0.10$~mag.
Thirty-two stars    were available for   a  calibration  of  $BVRI$ PL
relations; 25  stars were  used  for an alternative set  of  $BVRIJHK$
calibrations.  In the  following we  compare those multiwavelength  PL
relations with the Hipparcos sample of Galactic Cepheids, individually
corrected for  foreground  reddening  and  scaled to   their geometric
parallax distances.

We have collected from  the literature multiwavelength ($BVIJHK$) mean
magnitudes for as  many of the  Hipparcos-calibrating Cepheids as have
been  published (notably for  the infrared Wisniewski \& Johnson 1968,
Welch {\it et al.}  1984, Laney \& Stobie 1992 and reference therein).
These form rather  disjoint subsets.  After  eliminating the suspected
overtone pulsators   listed by FC97,  the total  available sample with
parallaxes drops from 26 to 20.  Of  these only 7 have mean magnitudes
published at  all  six    wavelengths,  while  10 and  13    Cepheids,
respectively have  either $BVIJK$ or $BVJK$ magnitudes  in common.  We
have  analyzed these   four   groups   of  stars independently,    but
self-consistently, in the following way.

Using the Hipparcos parallaxes and Galactic reddenings adopted by FC97
from Fernie, Kamper \& Seager (1993) scaled to the various wavelengths
using  the extinction  law of Cardelli,  Clayton  \& Mathis (1989), we
derived absolute  magnitudes for each of the  Cepheids  in each of the
observed  wavelengths.      (We   note that   these    corrections for
interstellar extinction are not inconsiderable, ranging up to 2~mag in
$B$ for  several stars).  The resulting PL  relations are shown in the
six  panels of Figure~2.  Error  bars are one-sigma uncertainties from
the quoted parallaxes.   Note   the highly  correlated nature   of the
individual data points about the fiducial lines. And too, we remind 
the reader that the computation of distances and their related errors  
from observed parallaxes is non-trivial (Brown {\it et al.} 1997), 
as distances are not linearly related 
to parallaxes, and parallax errors can subtly bias samples. A full 
treatment of this issue is beyond the scope of this paper, but we note that 
selection biases at least are minimized for stars having the smallest 
reported errors. As discussed by Brown {\it et al.}, given the 
true parallax distribution  the expected biases follow naturally;
however, corrections to the {\it observed } parallaxes require assumptions 
about the true distribution, and detailed modeling. Fortunately
for this application the 
Cepheid sample is not parallax-selected; the objects being chosen in 
advance based on their optical variability, periods and apparent magnitudes.

\placefigure{fig2}

The  differences  between   these individual  (trigonometric) absolute
magnitudes and the predicted $BVIJHK$ magnitudes derived from the mean
PL relations of  MF91  (solid lines in  Figure~2) are  each plotted in
Figure~3  against the  corresponding   B-band residual.   The  $(B-V)$
intrinsic color residuals are plotted against  the B-band residuals in
the upper right panel. The individual  residuals at a given wavelength
contain   random  contributions    from the parallax    uncertainties,
reddening  errors,  and  finally the   intrinsic (temperature-induced)
magnitude  residuals which reflect  the  finite  width of the  Cepheid
instability strip.  The observed residuals are however extremely large
(nearly 5~mag peak-to-peak) and are almost certainly dominated by the
(achromatic)  errors in the  parallaxes,  given the  strict unit-slope
correlation of the  mag-mag  residuals,  and the  total  lack of   any
correlation between the magnitude-color residuals (Figure~3).

\placefigure{fig3}

Wavelength-dependent   offsets  between   the   six  mean    solutions
independently will reflect (1) errors in the  adopted true distance to
the LMC (which set all of the  zero points in the MF91 multiwavelength
PL relation   calibrations),   (2) reddening errors  in    the adopted
extinction to the LMC sample of  calibrating Cepheids, and finally (3)
intrinsic differences  between  the  LMC  and Galactic  Cepheids,  for
example, due to metallicity.

Our  first solution  considers  the  largest  data  set (in   terms of
parallaxes) but the one that is most restricted in terms of wavelength
coverage:   it consists  of 19    Cepheids observed  in  $B$  and $V$.
Weighted  by the square of the  signal-to-noise ratio in the Hipparcos
parallax, the  residuals were summed and  averaged at each of  the two
wavelengths giving mean offsets  between  the LMC calibration and  the
Galactic  Cepheids.    The variance   in each mean   offset  was  then
calculated from the  average  of the squares  of these  same residuals
again inversely weighted  by  the variance in the  individually quoted
parallaxes.  The differences are $\Delta   B = +0.23~\pm0.35$~mag  and
$\Delta V   = +0.16~\pm0.28$~mag, in  the  sense that the  LMC Cepheid
calibration appears  to  be too faint   with respect  to  the Galactic
calibration.   (Further restricting the sample to  only those 12 stars
with    $\pi    /\sigma_{\pi} >    2.0$     changes   $\Delta B$    to
$+0.22~\pm0.24$~mag and $\Delta V$ to $+0.15~\pm0.17$mag.)

If the (statistically marginal, but apparently systematic) differences
in the $B$ and  $V$ solutions were to  be ascribed to reddening alone,
then the  Galactic data and the  LMC calibration can be  reconciled by
invoking an increase of $\Delta E(B-V) = 0.07$~mag in the adopted mean
reddening to the LMC Cepheid sample. This is consistent with a similar
suggestion  regarding  the LMC  Cepheid  calibration  made recently by
Bohm-Vitense (1997) based on different  data.  This reddening solution
has the consequence that it would also require the distance modulus of
the LMC to be revised {\it  downwards} by --0.06~mag to 18.44~mag; the
uncertainty on this offset being at least as  large as the uncertainty
in  the individual moduli  ($\pm0.3$~mag), depending on  the degree of
correlation in  those cumulative uncertainties.  This particular path,
of a  reddening  solution, cannot   be  considered definitive.   Other
possibilities are:  (1) the LMC  true  modulus should  be increased by
$(0.23  + 0.16)/2 =  +0.20$~mag, without any  change to the foreground
reddening, or (2) that  there are differential metallicity corrections
amounting to $-0.23$ and $-0.16$~mag  that need to  be applied at  the
$B$   and  $V$ wavelengths,   respectively.  Of   course  any suitably
contrived linear combination of the above  three effects could also be
invoked.  More constraints on the problem are obviously needed.

An alternative possibility is  that  some of the  wavelength-dependent
effects   seen in the comparison  of  Galactic (high metallicity) data
with  the  LMC  (lower metallicity)  data  could be    due to chemical
composition differences between the two  samples.  Taken at face value
the  dependence of the  apparent $V$  modulus  on metallicity would be
very   large, $\Delta  V/\Delta     [Fe/H]  = 0.16/0.15   =  1.1  (\pm
1.9)$~mag/dex, assuming that the full offset in $V$ noted in the above
comparison     is due  to metallicity,   and    adopting a metallicity
underabundance of  1.4$\times$   between   the  LMC  and    the  Solar
neighborhood (see, for   example, FW87).  However,  we  note that this
effect is basically indistinguishable from  reddening in its form  (as
evidenced by  our  first  set of    solutions), and that   the  offset
(whatever  its  origin)  when treated  as  reddening leads  to  a true
distance  modulus   for  the LMC   that   is unchanged,  from previous
assumptions, at  18.50~mag.   Given this  apparent  degeneracy between
reddening and metallicity, and the current  large uncertainties in the
parallaxes, assessing  the dependence on  metallicity from  these data
alone will remain problematic.

To obtain added  leverage on the  solution, moving to the infrared has
numerous  well known  advantages, as first   articulated in McGonegal,
McLaren, McAlary  \&  Madore (1982): reddening  effects  are known  to
decrease with wavelength, in a well defined and calibrated manner; and
simultaneously, metallicity effects are  also expected to  decrease in
amplitude with increased wavelength.

Our second solution is based on 13 Cepheids each having $BVHK$ data in
common.    This  four-color solution  gives  a  derived reddening {\it
increase}  for the LMC  Cepheid sample of $+0.04~\pm0.08$~mag, with no
formal offset in the derived $18.50~\pm0.13$~mag  true modulus for the
LMC.    Our next  approximation employs  10   Cepheids each now having
$BVIJK$ mean magnitudes. Here the formal solution for the true modulus
for the LMC is  $18.53~\pm~0.14$~mag, with a corresponding increase in
the mean reddening of  $+0.06~\pm~0.07$~mag. Finally, we have analyzed
a sample  of 7 Galactic Cepheids, each  having $BVIJHK$ photometry, to
obtain one last  solution:  $\Delta E(B-V) =   0.07~\pm~0.07$~mag with
$(m-M)_{LMC} = 18.57~\pm~0.11$~mag.  The   fit  to this final set   of
observations  is shown in  Figure~4;   the $\chi^2$ weighted  residual
fitting surface  being shown  as  an inset.   The individual  apparent
moduli discussed here, and their errors, are summarized in Table 1.

\placetable{tbl-1}

\placefigure{fig4}

Finally, if we  now adopt the metallicity   correction of $\Delta V  =
0.04$~mag advocated by  FC97 and assume that   the effects at  JHK are
negligible, (and  eliminate $B$ and  $I$ from the solution  given that
metallicity corrections for these filters are not well defined at this
time)   we  find   for  this    4-color  solution  $\Delta  E(B-V)   =
0.06~\pm~0.11$~mag  with $(m-M)_{LMC} =  18.57~\pm~0.11$~mag. This  is
virtually  indistinguishable  from the  full  $BVIJHK$ solution  given
above.

\section{Discussion and Conclusions}

We have used  the Hipparcos parallaxes of  nearby Galactic Cepheids to
explore corrections to the multiwavelength Period-Luminosity relations
for LMC Cepheids. The latter are based on an LMC  data set scaled to a
true distance modulus of 18.50~mag and an adopted foreground reddening
of  $E(B-V) =  0.10$~mag.  Although  the  current uncertainties in the
parallaxes are large and still dependent  upon the specific subsets of
the  Cepheids   chosen for  the  comparison, the   agreement  is good,
indicating  that to  within  $\pm$0.14~mag (or,  7\% in  distance) the
previously adopted  zero  point is  substantially correct.   Based  on
different  subsamples of data either having   $BV$, $BVJK$, $BVIJK$ or
$BVIJHK$ photometry, LMC moduli,  ranging from 18.44 to 18.57~mag  are
derived.  These results, summarized in Table 2,  differ from the value
of 18.70~mag of FC97, which are based soley on the reddening-corrected
V  photometry of Caldwell \&  Laney   (1991), externally adjusted  for
metallicity.  The Hipparcos data alone do not allow us to discriminate
between metallicity effects and the physically distinct possibility of
added reddening to the LMC.

\placetable{tbl-2}

To alleviate the  ambiguity posed by the  need to simultaneously solve
both for reddening and metallicity effects on the Cepheid distances we
are currently deriving OB-star reddenings  along the individual  lines
of  sight to  several  dozen LMC   Cepheids.  This will  allow  us  to
decouple the reddening determinations from metallicity effects, and go
beyond the use of a  single mean (foreground $+$ internal)   reddening 
for the LMC
calibrating Cepheid sample.   Preliminary reductions indicate that the
variance from field to field is large (ranging from $E(B-V) =$ 0.00 up
to  0.40~mag)   while  still indicating  that   an   average value  of
$<~E(B-V)~> ~=  0.10$~mag  is   appropriate for the    LMC calibrating
Cepheids.  Details  will be presented  in Madore, Freedman \& Pevunova
(1998 in preparation).

We  close   by   noting  that  at  least   three    other  very recent
determinations of the true modulus to  the LMC fall  on either side of
the value 18.50~mag  adopted by MF91 in  setting a zero point for  the
Cepheid distance scale.   Both  Reid (1997) and  Gratton  {\it et al.}
(1997)  derive  large LMC   moduli    (18.65 $\pm$~0.10,  and    18.63
$\pm$~0.06~mag,   respectively) using Hipparcos-based calibrations  of
the Galactic globular  cluster and  RR Lyrae  distance scale.  On  the
other   hand,  Gould  \&  Uza  (1997)   have  re-analyzed the SN~1987A
supernova ``light echo'' and derive {\it an upper limit} of $\mu_{LMC}
<$  ~18.37 $\pm0.04$~mag for the   LMC true distance modulus; although
they  note that if  the ring is  slightly elliptical ($b/a \sim$ 0.95)
this  upper limit  increases  to $<~18.44~\pm~0.05$~mag. 
A value of 18.56 $\pm$ 0.05~mag has been derived by Panagia 
{\it et al.} (1996) from the same data.   Until these differences are 
fully  understood and resolved, and given the remaining uncertainties in 
the Hipparcos Cepheid parallax data  we  prefer to adopt  a true distance 
modulus of 18.50~mag for the LMC, but  now bounded by an uncertainty 
of $\pm$~0.15~mag, defined to fully encompass the above range of
recently published  values. This value is consistent with other 
estimated distances to the LMC based on a wide variety of methods (for
a comprehensive modern review see Westerlund 1997). Viewed in 
that perspective the Hipparcos 
data confirm the  Cepheid distance scale at better than the $\pm$10\% 
level (95\% confidence).

\acknowledgments

BFM was supported in part by the Jet Propulsion Laboratory, California
Institute of Technology,  under the  sponsorship of  the  Astrophysics
Division of NASA's Office of Space Science and Applications and by the
NASA/IPAC Extragalactic   Database.    We acknowledge  and  appreciate
having had illuminating correspondence and discussions  with Drs.  
M.~Feast, A.~Gould, N.~Reid,  N.~Tanvir, C.~Turon and F.~van Leeuwen,  
in the course of preparing this paper.

\clearpage 

\figcaption[m1.eps]{Differential comparison  of    recently  published
V-band   PL  relations   (heavy  lines)   relative  to  the  Hipparcos
calibration   (thin    lines).  Plotted  is   the    difference [$V$ -
$V$(Hipparcos)]   versus log~P, in the  sense    that if Hipparcos  is
brighter the difference shown is positive. \label{fig1}}

\figcaption[m2.eps]{ Multiwavelength Period-Luminosity  relations  for
Cepheids with Hipparcos parallaxes, plotting all  stars that have data
available at  the particular wavelength,  as  noted in the upper  left
corner of each panel.   In each panel the  solid sloping line is not a
fit to the data, but rather it is  the published calibration of Madore
\& Freedman   (1991) flanked by thin   parallel lines representing the
2-$\sigma$ limits quoted  by them as being  the intrinsic width of the
instability strip at each wavelength. \label{fig2}}

\figcaption[m3.eps]{B-band   residuals    from   the   multiwavelength
Period-Luminosity relations in Figure  3 are sequentially plotted as a
function of  residuals from each  of the other   five PL relations and
(upper right panel) against the (B--V) color residuals. The total lack
of  correlation  in the latter  instance is  unexpected except  in the
limit  where  the residuals are   dominated by distance  errors in the
derived parallaxes. This latter situation is apparently the case given
the strong  (unit-slope) correlations of the residuals  in each of the
other panels, regardless of wavelength. \label{fig3}}

\figcaption[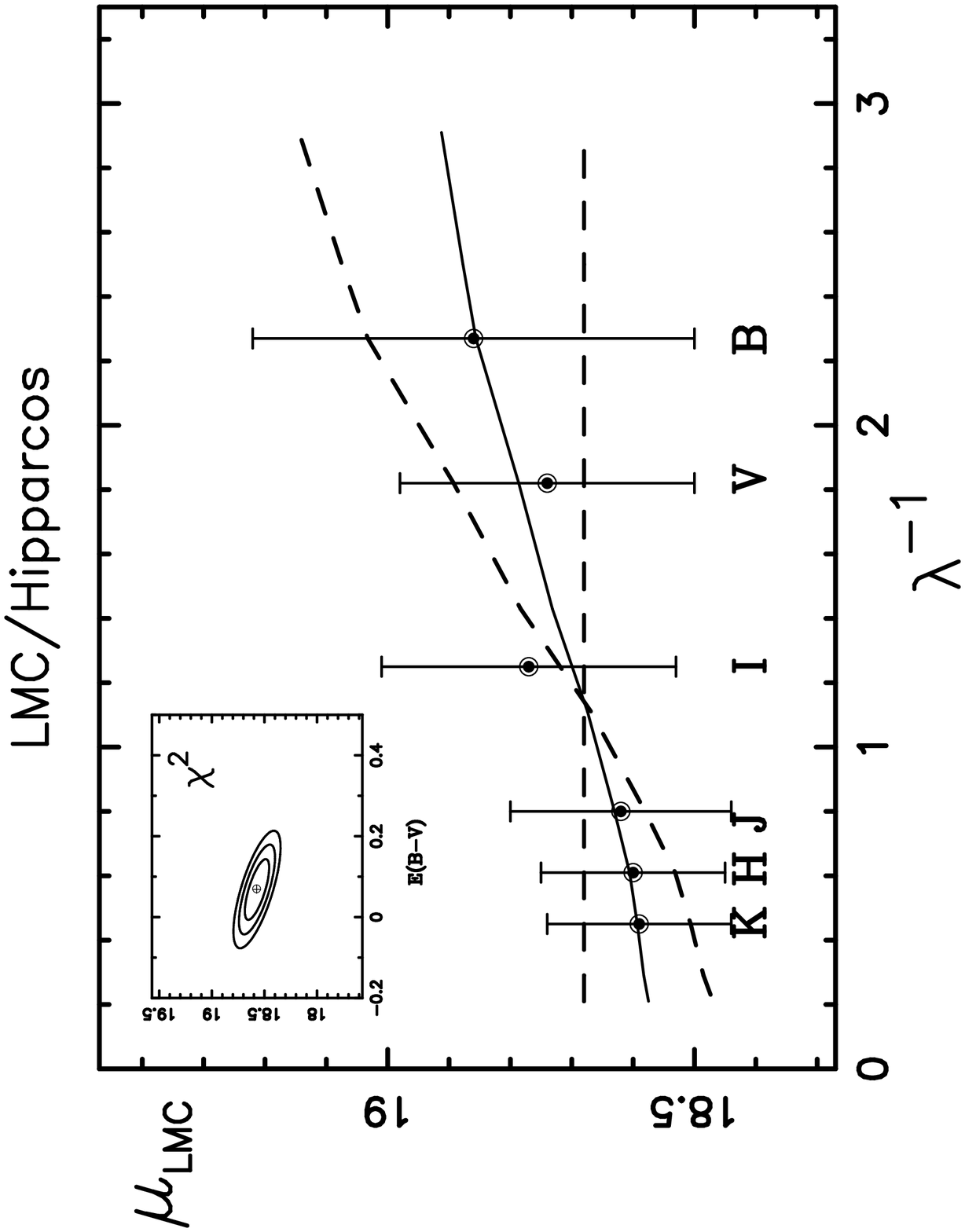]{ Apparent modulus plots  for LMC Cepheids observed
at $BVIJHK$ scaled to the Hipparcos zero point and using the published
multiwavelength PL solutions of  Madore \& Freedman (1991).  The solid
line is a weighted  $\chi^2$ fit of a  reddening line to the data; the
broken line indicates  the one-sigma limits  on that solution.   Inset
(top  left) shows  the $\chi^2$   surface indicating the  minimization
solution  for the  modulus and reddening   and the interdependence  of
their associated errors.\label{fig4}}

\clearpage 

\begin{deluxetable}{ccccccc}
\tablecolumns{7}
\tablewidth{0pc}
\tablecaption{Multiwavelength Moduli  for LMC Cepheids}
\tablehead{
\colhead{No. Stars}&\colhead{$\mu_B\pm \sigma$}&\colhead{$\mu_V\pm \sigma$}&\colhead{$\mu_I\pm \sigma$}& 
\colhead{$\mu_J\pm \sigma$}&\colhead{$\mu_H\pm \sigma$}&\colhead{$\mu_K\pm \sigma$}
}
\startdata
19&18.73$\pm$0.35&18.66$\pm$0.28&\nodata &\nodata&\nodata &\nodata\nl
13&18.71$\pm$0.36&18.64$\pm$0.24&\nodata&18.44$\pm$0.23&\nodata&18.54$\pm$0.13\nl
10&18.74$\pm$0.36&18.67$\pm$0.24&18.71$\pm$0.20&18.44$\pm$0.24&\nodata&18.57$\pm$0.14\nl
 7&18.86$\pm$0.36&18.74$\pm$0.24&18.77$\pm$0.24 &18.62$\pm$0.18&18.60$\pm$0.15&18.59$\pm$0.15\nl
\enddata
\end{deluxetable}

\begin{deluxetable}{clccc}
\tablecolumns{4}
\tablewidth{0pc}
\tablecaption{Multiwavelength Reddening Solutions}
\tablehead{
\colhead{}&\colhead{Filters} & \colhead{No. Stars}   & \colhead{$E(B-V) \pm \sigma$}   & \colhead{$\mu_{LMC} \pm \sigma$} 
} 
\startdata
&BV &19&0.17 $\pm$ \nodata &18.44 $\pm$ 0.35 \nl
&BVJK &13&0.14 $\pm$ 0.08&18.50 $\pm$ 0.13 \nl
&BVIJK &10&0.16 $\pm$ 0.07&18.53 $\pm$ 0.14 \nl
&BVIJHK &7&0.17 $\pm$ 0.07&18.57 $\pm$ 0.11 \nl
&V$_c$JHK &8&0.16 $\pm$ 0.11&18.57 $\pm$ 0.11 \nl
\enddata
\end{deluxetable}
\vskip3truein

\clearpage
\plotfiddle{m1.eps}{3in}{270}{70}{70}{-270}{250}
\vskip2.0truein
\clearpage
\plotfiddle{m2.eps}{3in}{270}{70}{70}{-270}{250}
\vskip2.0truein
\clearpage
\plotfiddle{m3.eps}{3in}{270}{70}{70}{-270}{250}
\vskip2.0truein
\clearpage
\plotfiddle{m4.eps}{3in}{270}{70}{70}{-270}{250}
\vskip2.0truein

\end{document}